\documentclass[%
 prd,
 reprint,
superscriptaddress,
preprintnumbers,
nofootinbib,
 amsmath,amssymb,
 aps,
]{revtex4-1}
\pdfoutput=1
\usepackage{hyperref}

\usepackage[textwidth=1.25cm,textsize=footnotesize,
]{todonotes}

\usepackage[utf8]{inputenc}

\newcommand{\eV}{\text{\,eV}}
\newcommand{\keV}{\text{\,keV}}

\newcommand{\be}{\begin{equation}}
\newcommand{\ee}{\end{equation}}
\newcommand{\ba}{\begin{eqnarray}}
\newcommand{\ea}{\end{eqnarray}}

\newcommand{\Zn}{\mathop\mathrm{Zn}}
\newcommand{\Cr}{\mathop\mathrm{Cr}}
\newcommand{\Ar}{\mathop\mathrm{Ar}}
\newcommand{\Ga}{\mathop\mathrm{Ga}}
\newcommand{\Ge}{\mathop\mathrm{Ge}}

\begin{document}

\preprint{INR-TH-2017-021}

\title{Revised neutrino-gallium cross section and 
 prospects of BEST in resolving the Gallium anomaly}
\author{Vladislav Barinov}
\email{barinov.vvl@gmail.com} 
\affiliation{Institute for Nuclear Research of the Russian Academy of Sciences,
  Moscow 117312, Russia}
\affiliation{Physics Department, Moscow State University, 
Vorobievy Gory, Moscow 119991, Russia}

\author{Bruce Cleveland}
\email{bclevela@snolab.ca}
\affiliation{SNOLAB, 1039
Regional Road 24 Lively ON, P3Y 1N2, Canada}

\author{Vladimir Gavrin}
\email{gavrin@inr.ru}
\affiliation{Institute for Nuclear Research of the Russian Academy of Sciences,
  Moscow 117312, Russia}

\author{Dmitry Gorbunov}
\email{gorby@ms2.inr.ac.ru}
\affiliation{Institute for Nuclear Research of the Russian Academy of Sciences,
  Moscow 117312, Russia}
\affiliation{Moscow Institute of Physics and Technology, 
  Dolgoprudny 141700, Russia}

\author{Tatiana Ibragimova}
\email{tvi@inr.ru}
\affiliation{Institute for Nuclear Research of the Russian Academy of Sciences,
  Moscow 117312, Russia}


\begin{abstract}
${\cal O}(1)$\!\eV\, sterile neutrino can be responsible for a number
of anomalous results of neutrino oscillation experiments. This
hypothesis may be tested at short base-line neutrino oscillation
experiments, several of which are either ongoing or under
construction.  Here we concentrate on the so-called Gallium anomaly,
found by SAGE and GALLEX experiments, and its foreseeable future
tests with BEST experiment at Baksan Neutrino Observatory. We start
with a revision of the neutrino-gallium cross section, that is
performed by utilizing the recent measurements of the nuclear final
state spectra. We accordingly correct the parameters of
Gallium anomaly and refine the BEST prospects in testing it and
searching for sterile neutrinos. We further evolve the previously
proposed idea to investigate the anomaly with ${}^{65}\!\Zn$ artificial
neutrino source as a next option available at BEST, 
and estimate its sensitivity to the sterile neutrino model parameters 
following the Bayesian approach. We show that after
  the two stages of operation BEST will make 5\,$\sigma$-discovery of
the  sterile neutrinos, if they are behind the Gallium anomaly. 
\end{abstract}

\maketitle

\section{Introduction}
\label{sec:Intro}

Sterile neutrinos are hypothetical massive Majorana fermions, singlets
with respect to the Standard Model (SM) gauge group, which have been
introduced to explain oscillations of the SM (or active) neutrinos via
mixing with them. There are no direct evidences for the
sterile neutrinos, unless one interprets the results of the
(anti)neutrino oscillation experiments, missing a fit by three active
neutrinos, as observations of ${\cal O}(1)$\ eV sterile neutrinos.
Though rather speculative, this interpretation encourages physicists
to put forward various
experimental proposals\,\cite{Abazajian:2012ys} to check this
hypothesis and hunt the sterile neutrinos.  One of such proposals,
Baksan Experiment on Sterile Transitions (BEST)
  \cite{Gavrin:2010qj,Abazajian:2012ys,Gavrin:2015aca,Barinov:2016znv,Gavrin:2017shq} is
a short base-line oscillation experiment aimed at searches/measurements
of disappearance of electron neutrino by capturing it on gallium,
\begin{equation}
	\nu_e + {}^{71}\!\Ga\rightarrow {\text e}^- + {}^{71}\!\Ge
	\label{Gallium-Germanium}
\end{equation}
Neutrinos come from an artificial source, which is supposed to be
${}^{51}\!\Cr$. It provides with direct testing the
Gallium anomaly \cite{Abdurashitov:1998ne,Abdurashitov:2005tb,Kaether:2010ag} 
which includes 4 measurements in total, and 3 out of 4 was performed
with ${}^{51}\!\Cr$ sources. 

In this paper we study the recently proposed idea
\cite{Gavrin:2017shq} to perform after ${}^{51}\!\Cr$-based experiment
the second
stage of the BEST operation with the
neutrino source based on the isotope ${}^{65}\!\Zn$. The main 
advantage of ${}^{65}\!\Zn$ with respect to ${}^{51}\!\Cr$
is higher availability. At
the same time,  
neutrino spectra of ${}^{51}\!\Cr$ and
${}^{65}\!\Zn$ are significantly different which allows us to achieve
more 'uniform' sensitivity to the sterile neutrino parameter space 
after BEST subsequent operation with the two artificial sources.  In
this case, to estimate the BEST sensitivity to sterile neutrino
parameters we calculate the cross section of the process
\eqref{Gallium-Germanium} at neutrino energies expected for the
${}^{65}\!\Zn$ source. To this end we use the computer program
\texttt{"speccros"} by John Bahcall which we adapt to account for
recent measurements of Refs.\cite{Frekers2011134,Frekers2013233}. We
revise the estimates of the cross section of process
\eqref{Gallium-Germanium} utilized by SAGE and GALLEX, and
consequently refine the parameters of sterile neutrino model favored
by the Gallium
anomaly\,\cite{Giunti:2010zu,Giunti:2012tn,Barinov:2016znv}. Then we
reestimate the prospects of testing the Gallium anomaly at BEST with
artificial source based on ${}^{51}\!\Cr$ isotope. Finally, we find
the sensitivity of BEST with ${}^{65}\!\Zn$ source to the sterile
neutrino model parameters. In accord with expectation, we observe that
running the subsequent experiments with ${}^{51}\!\Cr$ and
${}^{65}\!\Zn$ neutrino sources improves considerably the BEST
performance. In particular, it would allow to make 
  5\,$\sigma$-discovery and determine the sterile neutrino model
  parameters with 10\% accuracy.

The paper is organized in the following way. The neutrino-gallium
cross section is revisited in Sec.\,\ref{sec:cross-section}. In
particular, here we obtain formulas valid for ${}^{37}\!\Ar$,
${}^{51}\!\Cr$ and ${}^{65}\!\Zn$ sources. Sec.\,\ref{sec:BEST}
contains a sketch of BEST. In Sec.\,\ref{sec:Results} we apply the
obtained in previous sections results to refine the Gallium anomaly,
revise the BEST sensitivity with ${}^{51}\!\Cr$ source and investigate
BEST perspectives with ${}^{65}\!\Zn$ source in testing the Gallium
anomaly and searches for sterile neutrinos. We summarize in
Sec.\,\ref{sec:Summary}.

\section{Neutrino-gallium cross section}
\label{sec:cross-section}

The general formula for neutrino absorption cross section accounting
for nuclear transitions in reaction \eqref{Gallium-Germanium} can be
cast in the following form \cite{PhysRevC.56.3391}:
\begin{equation}\label{eq:sigma}
	\sigma = \sigma_0 \langle \omega_{e}^2G(Z,\omega_{e})\rangle\,,
\end{equation}
where expression in brackets is the dimensionless phase space factor
and $\sigma_0$ refers to the scale of the neutrino capture cross
section. 

Scale factor $\sigma_0$ can be written as \cite{BachallNA,Zuber127}
\begin{equation}
	\label{scale-factor}
	\sigma_0 = \frac{4 \pi^3\, \log \!2\, \alpha \hbar^3}{m_{e}^3c^4}\left(\frac{2J_f + 1}{2J_i +1}\right)\frac{Z}{ft_{1/2}(^{71}\!\Ge)}\,,
\end{equation}
where $\alpha$ is the fine-structure constant, $J_f$ is spin of the
final nuclear state, $J_i$ is spin of the initial nuclear state, $Z$ is
atomic number of the final nucleus, $ft_{1/2}(^{71}\!\Ge)$ is the
product of 
dimensionless phase-space factor $f$ for the kinematically allowed {\it
	electron capture}, the inverse process to the reaction
\eqref{Gallium-Germanium}, and $t_{1/2}(^{71}\!\Ge)$ is the half-life of
$^{71}\!\Ge$. This factor is defined as
\begin{equation}
  \label{ft-half}
	ft_{1/2}(^{71}\!\Ge) \equiv \frac{2\pi^3\,\log\!2\,\hbar^7}{m_{e}^5c^4}\frac{1}{(G_V^2|M_{i,f}|_{F}^2 + G_A^2|M_{i,f}|_{GT}^2)},
\end{equation}
where $ G_ {V}, G_ {A} $ are the vector and axial coupling constants
of nucleon, determined from the neutron decay \cite{PDG}, and $ | M_
{i,f} | _ {F} ^ 2 $, $ | M_ {i,f} | _ {GT} ^ 2 $ are the squares of the
transition matrix elements, which the vector current (Fermi
transitions) and the axial-vector current (Gamow--Teller
transitions) contribute to \cite{Wu,PhysRevC.74.054303}. These
allowed transitions are illustrated in Table\,\ref{tabl_transitions},
\begin{table}[htb!]	
\caption{Types of allowed transitions. S is total spin of
		the leptons. $ \Delta $L is change of the total angular
		momentum of the system. $ \Delta $P is change of parity of
		the system.
		\label{tabl_transitions}}	\begin{center}
		\begin{tabular}{c | c}
			\hline
			Fermi Transitions  \hspace{0.05cm} &  \hspace{0.05cm} Gamow--Teller Transitions \\	
			\hline
			\hline	
			\\
			$ \frac{1}{2}\uparrow_n $ $ \rightarrow $ $ \frac{1}{2}\uparrow_p $ + $  \frac{1}{2}\uparrow_e $ + $  \frac{1}{2}\downarrow_{\nu} $  \hspace{0.05cm} & \hspace{0.05cm} $ \frac{1}{2}\uparrow_n $ $ \rightarrow $ $ \frac{1}{2}\downarrow_p $ + $  \frac{1}{2}\uparrow_e $ + $  \frac{1}{2}\uparrow_{\nu} $ \\
			\\
			S = 0, $ \Delta $ L = 0, $ \Delta $$ \text{P} $ = 0 \hspace{0.05cm} &  \hspace{0.05cm} S = 1 , $ \Delta $L = $ \pm $ 1, $ \Delta $$ \text{P} $ = 0 \\	
			\hline
			\hline
		\end{tabular}
	\end{center}
	\end{table}
and the squared 
transition matrix elements read \cite{Wu},
\begin{align}
  \label{F-squared}
	|M_{i,f}|_{F}^2 & = |\langle f |\sum_{n=1}^{A}Q_{n}^{+}| i \rangle |^2\,,\\
	|M_{i,f}|_{GT}^2 & = \sum_{j=-1,0,1}| \langle  f
	||\sum_{n=1}^{A}Q_{n}^{+}\sigma_j| i \rangle |^2\,,
        \label{GT-squared}
\end{align}
where $ Q_ {n} ^ {+} $ is the transformation
operator of neutron into proton without a spin flip, and the
sum is taken over all nucleons in the nucleus; $2\times2$ spin matrices
$ \sigma_j $ are related to the 
Pauli matrices $\tau_i$ as follows
\begin{equation}
	\sigma_1 = \frac{1}{\sqrt{2}} \left(\tau_1 + i\tau_2 \right),\, \sigma_0 = \tau_3,\, \sigma_{-1} = \frac{1}{\sqrt{2}} \left(\tau_1 - i\tau_2 \right). 
\end{equation}
Summations in \eqref{F-squared},\,\eqref{GT-squared} go over the spin matrices for all possible
orientations of the angular momentum of the nucleon in the final
state, since the transition probability (due to invariance with respect
to rotations) should not depend on the magnetic
quantum number of the initial state.

Following the works of John Bahcall
\cite{PhysRevC.56.3391,RevModPhys.50.881,BachallNA}, we introduced in
\eqref{eq:sigma} 
the value of $ \langle
\omega_ {e} ^ 2G (Z, \omega_ {e}) \rangle $, where $ G (Z, \omega_
{e}) \equiv p_eF(Z, \omega_ {e}) / 2 \pi \alpha Z \omega_e $, is 
dimensionless phase-space factor averaged over the electron
energies. The explicit expression is given by 
formula
\begin{equation}\label{eq:factor}
	\langle \omega_{e}^2G(Z,\omega_{e})\rangle \equiv 
	\frac{\int_{\omega_{e}^{min}}^{{\omega_{e}^{max}}}\omega_{e}p_{e}F(Z,
		\omega_{e})\phi(q_\nu)d\omega_{e}}{2\pi\alpha Z \int_{0}^{{q_{\nu}^{max}}}\phi(q_\nu)dq_\nu}\,,
\end{equation} 
where $ \phi (q _{\nu}) $ is the neutrino energy
distribution function, $ q_{\nu} = E_{\nu}/m_ec^2 $ is the
dimensionless neutrino energy, $ \omega_e \equiv E /m_ec^2 $, $ p_e =
p/m_ec $ are the dimensionless energy and momentum of the electron. The
integrals in \eqref{eq:factor} are taken over the whole spectrum of
electrons, which energy can be expressed as 
\begin{equation}\label{eq:E_Q}
	E = E_{\nu} + [M(A,Z-1)-M(A,Z)]c^2 + m_ec^2 -  \langle E_{ex}
        \rangle - V_{0},
\end{equation}
where $ E _{\nu} $ is energy of the incoming neutrino,
$ \langle E_{ex} \rangle $ is average
excitation energy of the produced nucleus, $ V_{0} $
is a correction \cite{BAHCALL196610} for smaller average binding energy of electron inside the
nucleus with respect to that outside,   
and term in parenthesis is the atomic mass difference between initial $ M (A, Z-1)$ and final $ M (A, Z)$ atomic masses.

Quantity $F(Z, \omega_{e})$, which enters into formula
\eqref{eq:factor}, accounts for the non-planewave structure
of the electron wave-function and is  
closely related to the Fermi function\,\cite{Konopinski}, that is the
ratio of electron squared wave functions calculated with and without
the Coulomb potential, 
\begin{equation}\label{eq:fermi}
	F(Z,E,r) = 2(1+\gamma_0)(2pr/\hbar)^{2(\gamma_0 -1)}e^{\pi\nu}\frac{|\Gamma(\gamma_0 + i\nu)|^2}{[\Gamma(2\gamma_0 + 1)]^2}\,.
\end{equation}
Here we introduced $ \gamma_0 \equiv [1-(\alpha Z)^2]^{1/2} $, $ \nu
\equiv 
\alpha ZE/p_ec $, 
and $ r $ denotes distance from the nucleus center to the electron.
According to \cite{BachallNA} expression (\ref{eq:fermi}) must be
averaged over the entire finite volume $V$ of the nucleus of radius $ R $,
that reveals 
\begin{equation}
	\label{eq: averfermi}
	\begin{split}
		F(Z, \omega_{e})&= \frac {1}{V}\int_{0}^{R} F (Z, \omega_{e}, r) dV \\
		&= \left [\frac {3} {2 \gamma_0 + 1} \right] F(Z, \omega_{e}; r = R)\,.
	\end{split}
\end{equation}
The resulting correction reflects the fact that electron capture can
occur at any point inside the nucleus.
There are also corrections \cite{BAHCALL196610} to $F(Z, \omega_{e})$  
due to shielding of the Coulomb potential inside the nucleus.
We take them into account, but find them small, at the level of
percent for the set of interesting neutrino energies. 

The review presented above in this Section concerns only the
allowed nuclear transitions. The question of the
contribution of the excited states of the nucleus to the total
neutrino absorption cross section is discussed below.

In paper \cite{HATA1995422} Hata and Haxton have shown that the
contribution of excited states to the total neutrino absorption cross
section on $^{71}\!\Ga$ can be written as 
\begin{equation}\label{eq:ex_st}
	\sigma = \sigma_{g.s.} \left[ 1 +
	\frac{\sum_{E_{x}}\lambda_{E_{x}}B(GT)_{E_{x}}}{B(GT)_{g.s.}}
	\right]\,.
\end{equation}
Here $\sigma_{g.s.}$ is the neutrino absorption cross section
associated with gallium $^{71}\!\Ga$ transition to the ground
state of germanium $^{71}\!\Ge$, which is given by
eq.\,\eqref{eq:sigma},
the coefficients $\lambda_{E_{x}}$ are the 
phase space factors for these transitions normalized to the ground-state
phase space factor \cite{Giunti:2012tn}.
These coefficients can be calculated from
eq.\,\eqref{eq:factor} by making use of
the program \texttt{"speccros"} written by
John Bahcall, $B(GT)_{g.s.}$ is the square of the Gamow--Teller
transition matrix
element to the ground state (see Table\,\ref{tabl_transitions}),
and $B(GT)_{E_x}$ are the squared 
matrix elements responsible for transitions to excited
states with energies $E_x$ \cite{PhysRevC.91.034608} measured from the
ground state. 

\begin{figure}[!htb]
	\centering
	\includegraphics[width=1.0\linewidth, height=0.25\textheight]{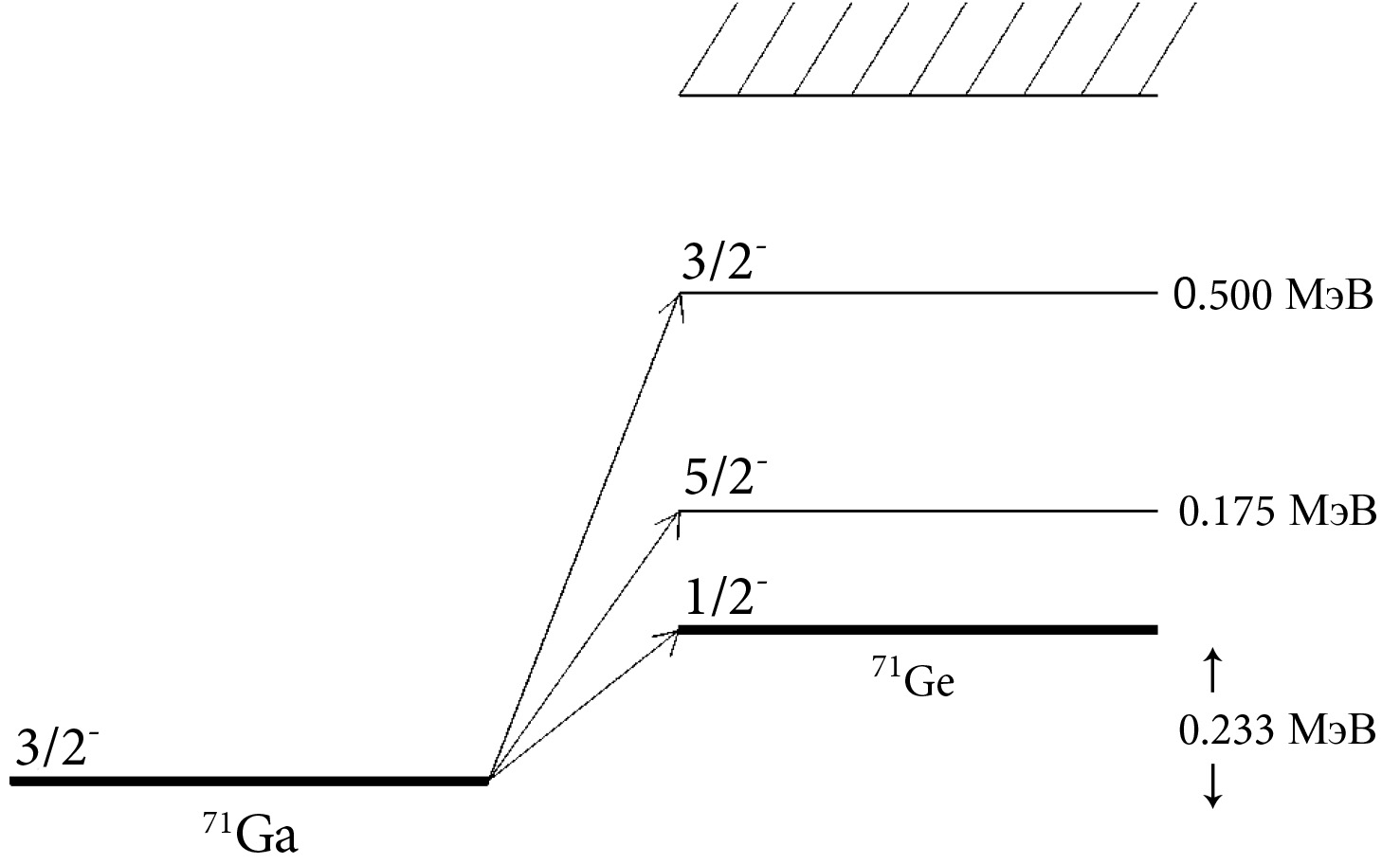}
	\caption{Scheme of the $^{71}\!\Ga$ $\rightarrow$ $^{71}\!\Ge$
          transitions
		induced by electron neutrinos emitted in weak decays of
		$^{51}\!\Cr$ and $^{37}\!\Ar$.\label{fig:Ga_d}}
\end{figure}

The gallium decay scheme, depicted in Fig.\,\ref{fig:Ga_d},
presents transitions to excited
states with excitation energies $E_x$ of 175\,keV and 500\,keV,
relevant for artificial sources of neutrinos based on radioactive
isotopes $^{51}\!\Cr$ \cite{Abdurashitov:1998ne}
 and $^{37}\!\Ar$ \cite{Abdurashitov:2005tb}.

However, for artificial neutrino source $^{65}\!\Zn$ \cite{LNE}, the
higher energy levels get excited in the process
\eqref{Gallium-Germanium} and their contribution to the total cross
section is significant, $\sim 20\!-\!30\%$. The coefficients $\lambda_{E_{x}}$ for these
transitions for $^{65}\!\Zn$ are $ \lambda_{175} = 0.7969$, $
\lambda_{500} = 0.4791$, $ \lambda_{708} = 0.3145$, $ \lambda_{808} =
0.2466$, $ \lambda_{1096} = 0.0934$, the squared transition matrix
elements $B(GT)_{E_{x}}$ corresponding to these energies are given in
\cite{PhysRevC.91.034608}.

Based on the new results of measuring the threshold energy of the
gallium transition to the ground state of germanium
\cite{Frekers2013233}, 
\begin{equation}
	\label{new-Q}
	Q = 233.5 \pm
	1.2\,\keV
\end{equation} 
and using the half-life of
$^{71}\!\Ge$ ($t_{1/2}(^{71}\!\Ge) = 11.43 \pm 0.03 $\,d)
\cite{PhysRevC.31.666}, we calculate $\log ft_{1/2}$ 
using the $ft$-calculator \cite{NNDC}, 
\begin{equation}\label{eq:logft}
	\log ft_{1/2}( ^{71}\!\Ge) = 4.353 \pm 0.005\,.
\end{equation}
We exploit \eqref{eq:logft} further to calculate the Gamow--Teller
transition matrix element $B(GT)_{g.s.}$, which can be written as
\cite{BachallNA,Giunti:2012tn}, cf.\,\eqref{ft-half}, 
\begin{equation}
	B(GT)_{g.s.}\! =\! \left[\! \dfrac{2J_{f} + 1}{2J_{i}+1} \!\right]\!
	\dfrac{2\pi^3\,\log \!2\,\hbar^7}{G_{F}^2 |V_{ud}|^2 m_{e}^5c^4 g_{A}^2 ft_{1/2}( ^{71}\!\Ge)},
\end{equation}
where $G_F$ is the Fermi constant, $V_{ud}$ is the element of the
Cabibbo--Kobayashi--Maskawa mixing matrix \cite{PDG}, $ g_{A} =
-1.2723(23)$ is the axial coupling constant \cite{PDG} and $ G_{A}^2 =
G_{F}^2 |V_{ud}| ^2g_{A}^2$. Numerically one finds   
\begin{equation}\label{eq:bgtgs}
	B(GT)_{g.s.}=0.086 \pm 0.001\,,
\end{equation}
while from eqs.\,\eqref{scale-factor} and \eqref{eq:logft} 
\begin{equation}\label{eq:s0my}
	\sigma_0 = (8.6 \pm 0.1) \times 10^{-46}  \text {cm}^2.
\end{equation}
While the central value of \eqref{eq:s0my} is fully consistent
with previous estimate\,\cite{PhysRevC.56.3391}:
\begin{equation}\label{eq:s0B}
	\sigma_ {0}^{Bahcall} = (8.611 \pm 0.011) \times 10 ^ {-46}  \text {cm}^2
\end{equation}
the uncertainty saturated by that of \eqref{new-Q} is significantly
larger. It happened because the value \eqref{eq:s0my} was obtained from
analysis of the new data \cite{Frekers2013233}. 
We utilize the new estimate of the threshold energy of
the gallium transition to the ground state of germanium \eqref{new-Q},
in contrast to the old value
$Q=232.69 \pm 0.15$\,keV used previously in \cite{PhysRevC.56.3391}.  
We use the most recent value \eqref{new-Q} and hence
\eqref{eq:bgtgs}, which are consistent with previous results, while
their errors do not dominate the uncertainties of our estimates of the
neutrino-capture cross sections.

Further, for each spectral line of the artificial sources $^{51}\!\Cr$,
$^{37}\!\Ar$ and $^{65}\!\Zn$ presented\footnote{The lowest line of
  $^{65}\!\Zn$ with close to threshold energy\,\eqref{new-Q} $E=0.236$\,keV
  is kinematically forbidden to produce
  electron (after account of somewhat lower electromagnetic binding
  energy $V_0$ of electron
  inside the nucleus with respect to that outside,
  see eq.\,\eqref{eq:E_Q}).} in Table\,\ref{Tab:lines},
\begin{table}[htb!]	
\caption{Neutrino energy spectra--energies $E_\nu$ and neutrino
  fractions $f_{E_\nu}$-- and corresponding neutrino capture cross
  section on gallium for the set of artificial sources under
  consideration. 
		\label{Tab:lines}}	\begin{center}
		\begin{tabular}{c|c|c|c}
			\hline
                        isotope &			  $E_\nu$, MeV
                     & $f_{E_\nu}$,\,\% & $\sigma(E_\nu)$, $10^{-46}$\,cm$^2$\\	
			\hline\hline
$^{51}\!\Cr$ 
                        \\ &0.752
                        & 8.49(1) & $63.22\pm 1.40$
			\\ &0.747
                        & 81.63(1) & $62.58 \pm1.39$
			\\ &0.432
                        & 0.93(1) & $27.14 \pm 0.52$ 
	                \\ &0.427
                        & 8.95(1) & $26.72 \pm 0.51$
			\\\hline
$^{37}\!\Ar$ 
                        \\ &0.813
                        & 9.80(1) & $71.63\pm 1.62$ 
			\\ &0.811
                        & 90.20(1) & $71.35 \pm 1.61$
\\ \hline
$^{65}\!\Zn$
\\ &1.352
& 48.35(11) & $181.5\pm4.2$ 
			\\\hline			
		\end{tabular}
	\end{center}
	\end{table}
the values of $\sigma_{g.s.}$ and $\lambda_{E_x}$ entering
\eqref{eq:ex_st} are 
calculated from \eqref{eq:sigma} and data
\cite{PhysRevC.91.034608,Frekers2013233} by making use of the program
\texttt{"speccros"}. 
Subsequently, for each neutrino energy the neutrino capture cross
section is obtained including contributions of the kinematically allowed excited
states, see Table\,\ref{Tab:lines}. Then the total neutrino absorption cross sections 
for each artificial source are obtained
by summing over all energies weighted with the corresponding relative fractions,
\[
\sigma=\sum_{E_\nu} \sigma(E_\nu) f_{E_\nu}
\]
The results are as follows
\begin{align}\label{Chrome-cross}
	\sigma(^{51}\!\Cr)&=  (59.10 \pm 1.14) \times 10^{-46}  \text{cm}^2,\\
	\label{Argon-cross}
        \sigma(^{37}\!\Ar) &= (71.38 \pm 1.46) \times 10^{-46}
        \text{cm}^2,\\
        \label{Zinc-cross}
	\sigma(^{65}\!\Zn) &= ( 87.76 \pm 2.03 ) \times 10^{-46}
        \text{cm}^2.        
\end{align}
We use these estimates in the following Sections.

\section{Sketch of BEST}
\label{sec:BEST}

The BEST experiment is described in detail in
Ref.\,\cite{Barinov:2016znv}. Here we merely recall the general idea of this
experiment.

\begin{figure}[htb!]
	\centering
	\includegraphics[width=0.85\linewidth, height=0.3\textheight]{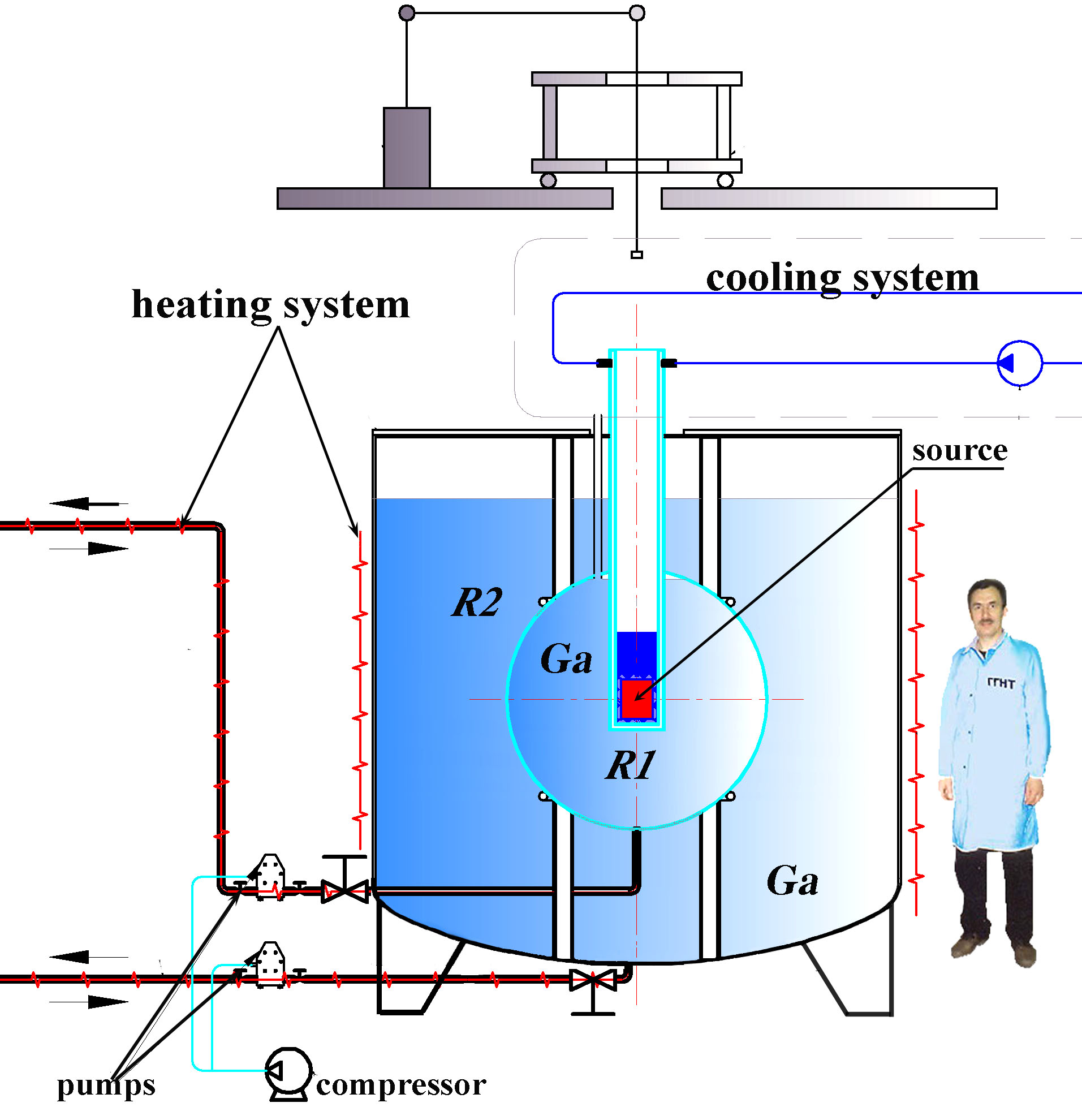}
	\caption{The BEST layout; vessel sizes are $R_1$ = 0.66\,m,
		$R_2 =$ 1.096\,m. 
		\label{fig:BEST-layuot}}
\end{figure}

The experimental setup consists of two concentric zones filled with
liquid gallium. The first zone is a sphere of radius $R_1$ = 0.66\,m,
in the center of which there is an artificial neutrino source about
0.1\,m in size. Such a size makes it possible to place in the center
of the first zone a source of neutrinos $^{51}\!\Cr$ with activity
3\,MCi. The second zone is a cylinder of radius $R_2$ = 1.096\,m and
height $2\times R_2$. The image of the experimental setup is shown in
Fig.\,\ref{fig:BEST-layuot}.
The liquid gallium is irradiated by a neutrino flux from an artificial
source. As a result of reaction \eqref{Gallium-Germanium}, germanium
atoms are formed, which are then chemically extracted from the
zones. Possible transitions to sterile neutrinos
would affect the neutrino flux. Hence the numbers of extracted atoms
are sensitive to the presence of light sterile neutrinos.

The total mass of gallium is 50 tons. The original
proposal\,\cite{Gavrin:2010qj,Abazajian:2012ys} suggests to exploit
the isotope $^{51}\!\Cr$ as the artificial neutrino source with
radioactivity of about 3\,MCi. At the same time other candidates may
be considered, and one of the most promising is $^{65}\!\Zn$
\cite{Gavrin:2017shq}. It provides different neutrino spectrum giving
the opportunity to test somewhat different region of sterile neutrino
parameter space.  Also the half-life of $^{65}\!\Zn$
  is longer (244 d compared to 27 d for $^{51}\!\Cr$), thus giving
  more time to make longer measurements with the sufficient activity
  of the source.  However, the artificial source $^{65}\!\Zn$ of the
same activity has a noticeably larger size than the source
$^{51}\!\Cr$, which reduces the oscillation signal after averaging over
the source volume.  This must be avoided, and a special investigation
is required to find the reliable technical solution and optimize the
source volume. For the present study we take as a realistic option to adopt
the smaller $^{65}\!\Zn$ source with activity of about 1\,MCi, which
will be acceptably compact.  The volume occupied by the source within
the first zone will increase slightly, but this will not negatively
affect the isotropy of target irradiation.  Likewise, with such
activity it will be possible to keep sufficiently high homogeneity of
the zinc source. Finally, the lower power of the source is partly
compensated by larger cross section \eqref{Zinc-cross}.
Although the predicted production rate from the
  $^{65}\!\Zn$ source with activity of about 1 MCi is about two times
  smaller compared to the 3MCi $^{51}\!\Cr$ source, nevertheless the
  expected number of germanium atoms to be extracted from the vessels
  are still sufficiently large with respect to the solar background. 
The statistical errors grow insufficiently and the total uncertainty
of the extraction is dominated by systematics, which we expect to be
the same as in case of $^{51}\!\Cr$ source.

\section{Revision of the Gallium anomaly and searches at BEST}
\label{sec:Results}

For the revision of the results for neutrino absorption cross sections we 
begin with discussion of uncertainties.

The main contribution to the uncertainty of the neutrino absorption
cross section is associated with corrections from the excited
states.  To calculate the
uncertainty of neutrino cross section, the results of
\cite{PhysRevC.91.034608,Frekers2013233}, as well as the known
uncertainty of $\sigma_0$ are accounted for. Assuming the measurements
of $B(GT)$ for different energy levels to be independent, we calculate
the overall error for each spectral line of the artificial sources as
the square root of the sum of the squared standard deviations of all
values entering \eqref{eq:ex_st}.

The obtained values of the cross sections for $^{51}\!\Cr$ and $^{37}\!\Ar$
and their relative
uncertainties deviate insignificantly from the previous study in 
\cite{Giunti:2012tn}. However, we take different value of 
the energy of the gallium transition to the ground state of germanium
\cite{Frekers2013233}, as well as another value of the
transition matrix element to the ground state \eqref{eq:bgtgs}.
We find the uncertainty of the cross sections to be about two
percent, while earlier for the BEST experiment
the uncertainty of +3.6\,\%/-2.8\,\% 
\cite{PhysRevC.56.3391} has been adopted.

It is worth noting that the measurement of the
threshold energy of the gallium-germanium transition does not contain
unknown uncertainties in the nuclear structure, which could explain
the anomalous results of the SAGE \cite{Abdurashitov:1998ne,Abdurashitov:2005tb} and GALLEX
\cite{Kaether:2010ag} experiments. This result was further discussed in
Ref.\,\cite{Frekers2013233}.

The results obtained in Section \ref{sec:cross-section} imply that
despite the fact that we applied new value of the
threshold energy of the gallium transition into the ground state of
germanium, than previously done, and despite the utilization of the
recent measurements of the transitions matrix elements
\cite{PhysRevC.91.034608}, the central values and
  their uncertainties have not changed much, in comparison with the
  values presented in \cite{Giunti:2012tn}. 
  The refined values of
 the ratios of observed-to-expected number of events $R$ in gallium
  experiments (gallium anomaly),  
  which we represent in this paper, see
  Table\,\ref{Table-3}, 
  almost completely
  coincide with the values presented in \cite{Giunti:2012tn}.

\begin{table}[h!]
	\caption{Values of the magnitudes of the
            gallium anomaly, obtained on the basis of refined data on
            the neutrino absorption cross section, using the value of
            $Q = 233.5 \pm 1.2 $ keV, the transition matrix element
            to the ground state $BGT_{g.s.} = 0.086 \pm 0.001$ and
            the transition matrix elements to excited states
            taken from Ref.\,\cite{PhysRevC.91.034608}.}
\label{Table-3}
	\begin{center}
		\begin{tabular}{c c c c c c}
			\hline
			 & SAGE 1 & SAGE 2 & GALLEX 1 & GALLEX 2 & AVE \\
			\hline
			$R$ & $0.93^{+ 0.12}_{- 0.12}$  & $0.77^{+ 0.09}_{- 0.08}$ & $0.93^{+ 0.11}_{- 0.11}$ & $0.80^{+ 0.11}_{- 0.11}$ & $0.84^{+ 0.05}_{- 0.05}$ \\
			\hline		
		\end{tabular}
	\end{center}
\end{table}

Thus, taking into account the refined value of the neutrino absorption
cross section on gallium found in this paper, the resulting error of
the experiment BEST \cite{Gavrin:2015aca} for the source $^{51}\!\Cr$
is 4.9\,\% for each of the zones and 4.2\,\% for the total target,
instead of 5.5\,\% and 4.8\,\%, respectively.  For the artificial
neutrino source $^{65}\!\Zn$ with activity of 1 MCi in the BEST
experiment, the resulting errors will be the same as for the 3\,MCi
$^{51}\!\Cr$ source if the irradiation plan with the $^{65}\!\Zn$
source is identical to that presented in Ref.\,\cite{Gavrin:2015aca}.

\begin{figure}[htb!]
	\centering
	\includegraphics[width=0.9\linewidth, height=0.3\textheight]{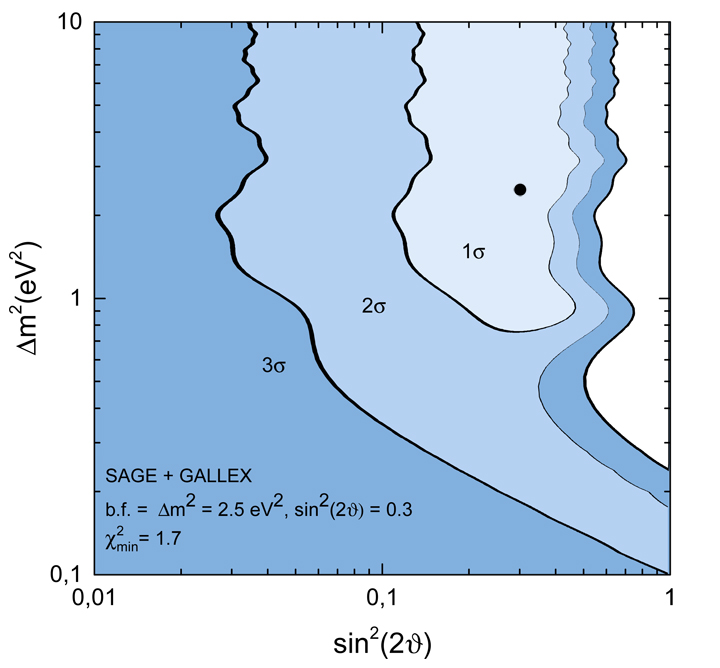}
	\caption{Allowed regions of oscillation parameters obtained by combining the results of SAGE + GALLEX using the refined data presented above.
		\label{fig:SG_NS_NR}}
\end{figure}

\begin{figure}[htb!]
	\centering
	\includegraphics[width=0.9\linewidth, height=0.3\textheight]{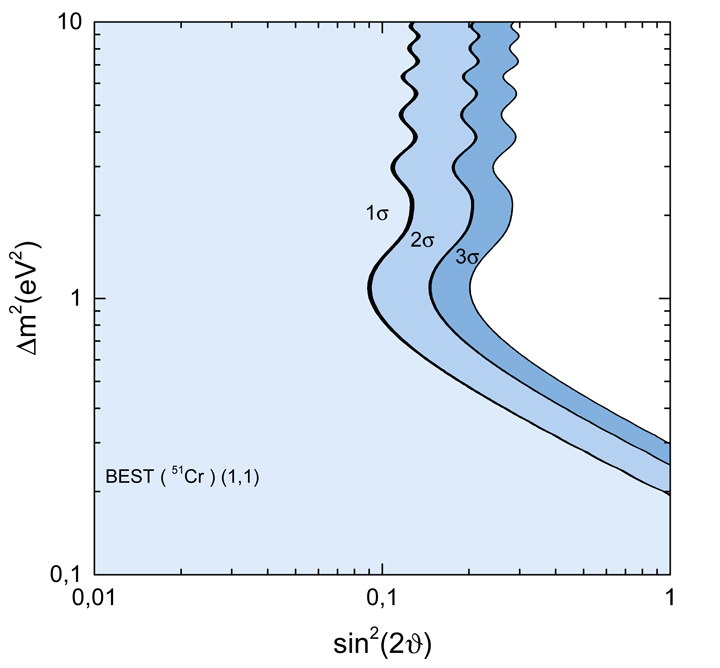}
	\caption{Allowed regions of oscillation parameters
		in case the BEST experiment does not find
		any anomalies: the ratios $R$ of observed-to-expected without sterile
		neutrinos germanium atoms in both vessels are consistent with unity, $(1,1)$.
		\label{fig:BCr_11_NS_NR}}
\end{figure}

\begin{figure}[htb!]
	\centering
	\includegraphics[width=0.9\linewidth, height=0.3\textheight]{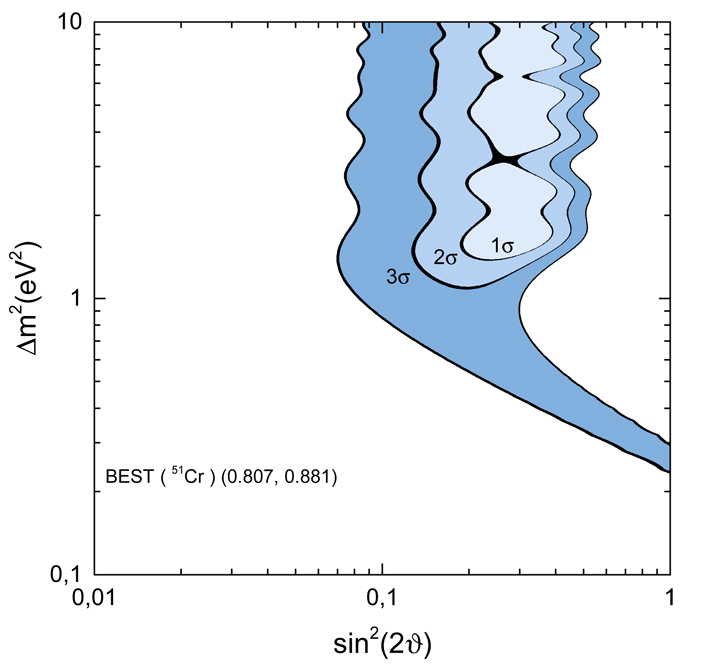}
	\caption{Allowed regions of oscillations parameters
		if the result of the BEST experiment corresponds
		to the best fit point for combining the SAGE + GALLEX. The numbers
		in parentheses indicate the most probable ratios $R$ of
		observed-to-expected without sterile neutrinos germanium atoms in
		the two vessels.
		\label{fig:BCr_BF_NS_NR}}
\end{figure}

\begin{figure}[htb!]
	\centering
	\includegraphics[width=0.9\linewidth, height=0.3\textheight]{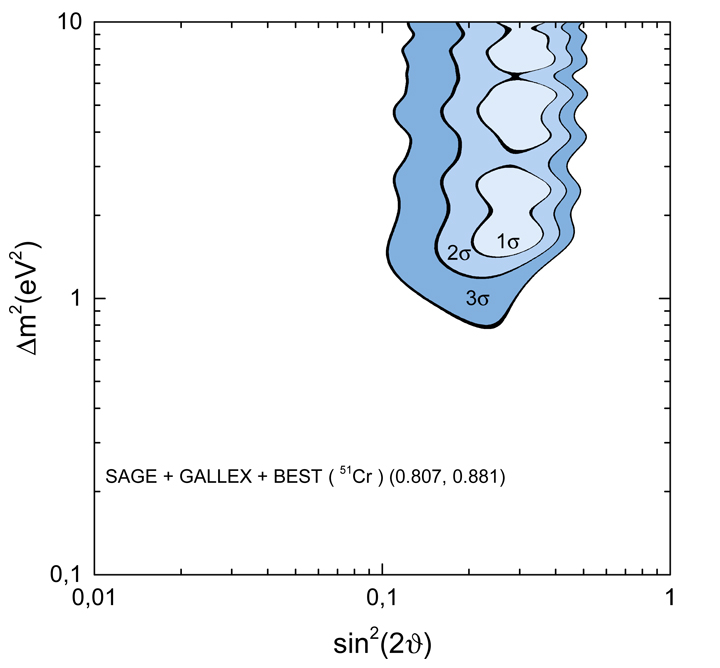}
	\caption{Allowed regions of oscillation parameters, when the results
		of SAGE + GALLEX are combined with the result of BEST for $^{51}
		\!\Cr$ source consistent with the SAGE+GALLEX best fit point.
		\label{fig:SGBBF_NS_NR}}
\end{figure}

The anomalous lack of neutrinos presented in Table\,\ref{Table-3} can be
  explained by oscillations of electron neutrinos into sterile
  partners\,\cite{Laveder:2007zz}.     
The combined results of SAGE and GALLEX, obtained on the basis of
refined data, are presented in Fig.\,\ref{fig:SG_NS_NR}. The result
shown in Fig.\,\ref{fig:SG_NS_NR} shows that the best
  fit values $\Delta m^2 =
2.5$ $ \text{eV}^2 $ and $ \sin^2(2\vartheta) = 0.3 $ are slightly
different (by about 10\%) from those presented in \cite{Barinov:2016znv}. The refined
regions of the neutrino oscillation parameters to be tested at the
BEST \cite{Gavrin:2015aca} experiment with the artificial source
$^{51} \!\Cr$ are given in Figs.\,\ref{fig:BCr_BF_NS_NR} and
\ref{fig:BCr_11_NS_NR}.
They are found by applying the formulas from
\cite{Barinov:2016znv}. 
Assuming the BEST with source $^{51}\!\Cr$ fully confirms the
anomaly, the most favorable regions (all data of the three experiments
are included) of sterile neutrino model
parameter space are presented in Fig.\,\ref{fig:SGBBF_NS_NR}. 
Comparing these plots with similar ones in 
  Ref.\,\cite{Barinov:2016znv} one can conclude that after revision of
  the neutrino capture cross section all signal regions become more
  compact, hence the sensitivity of BEST to the sterile neutrino
  model certainly increases. 

\begin{figure}[htb!]
	\centering
	\includegraphics[width=0.9\linewidth, height=0.3\textheight]{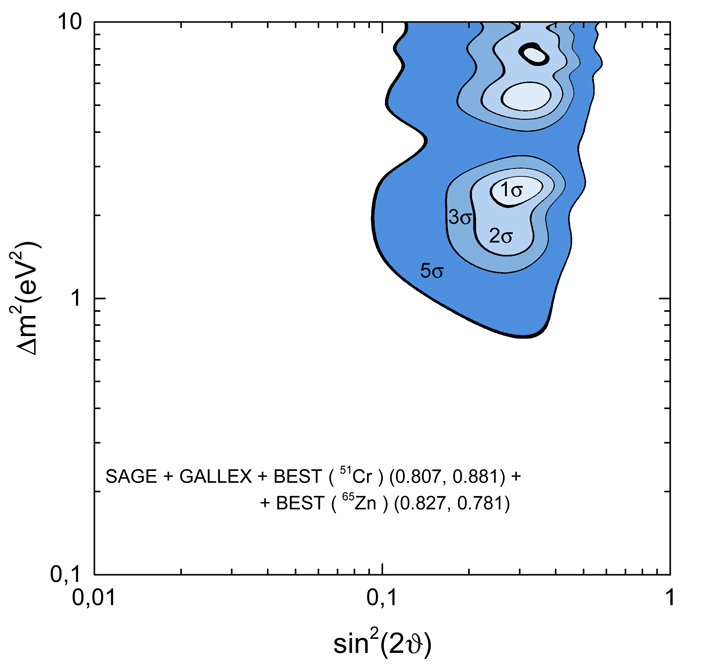}
	\caption{Allowed regions of oscillation parameters, built on the basis of new data, in the case of combining the results of SAGE + GALLEX with the result of BEST for two sources ($^{51}\!\Cr$ and $^{65}\!\Zn$), which corresponds to the best fit point.
		\label{fig:SGB1B2_BF_NS_NR}}
\end{figure}

\begin{figure}[htb!]
	\centering
	\includegraphics[width=0.9\linewidth, height=0.3\textheight]{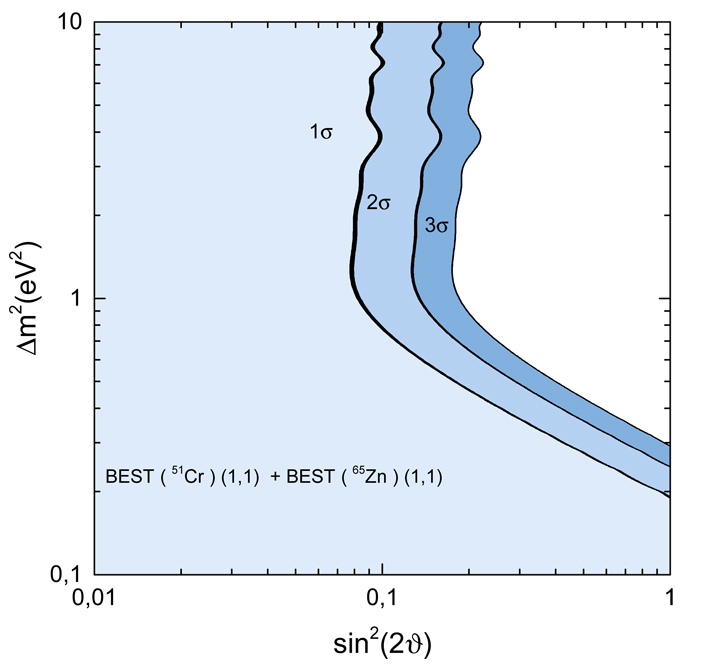}
	\caption{Allowed regions of oscillation parameters
		in case the BEST experiment does not find
		any anomalies after two runs:
		the ratios $R$ of observed-to-expected without sterile
		neutrinos germanium atoms in both vessels for both sources
		are consistent with unity, $(1,1)$.
		\label{fig:BZn_Cr_11_NS_NR}}
\end{figure}

To illustrate the power of the source $^{65}\!\Zn$ in further
testing the sterile neutrino hypothesis, we present in
Fig.\,\ref{fig:SGB1B2_BF_NS_NR}
the anomaly-favored region after the
second run of BEST operating with the source $^{65}\!\Zn$. 
The sensitivity of the second run is estimated in
exactly the same way as has been done in \cite{Barinov:2016znv} for
the source $^{51}\!\Cr$.  For the favored by gallium anomaly best fit values of the
  sterile neutrino model the expected signal rates in the two vessels
  of BEST correspond to ratios $R$=(0.827,0.781).   
One clearly observes from Figs.\,\ref{fig:SGBBF_NS_NR}
and \ref{fig:SGB1B2_BF_NS_NR}  
  the significant
improvement in the sensitivity after the combined analysis of the two
runs (assuming both confirm the Gallium anomaly). Finally, if both
runs find no hint of sterile neutrinos, the exclusion region will
expand with respect to that in Fig.\,\ref{fig:BCr_11_NS_NR}, and it is
presented in Fig.\,\ref{fig:BZn_Cr_11_NS_NR}.

\section{Summary}
\label{sec:Summary}

In this work updated data \cite{PhysRevC.91.034608,Frekers2013233} on
neutrino absorption cross section on gallium and the
program \texttt {"speccros"} are used to refine the neutrino
absorption cross section, which is done for $^{71}\!\Ga$ and 
neutrino sources $^{51}\!\Cr$, $^{37}\!\Ar$ and
$^{65}\!\Zn$.

The results obtained for the sources $^{51}\!\Cr$ and
$^{37}\!\Ar$ agree with the estimates presented in \cite
{Giunti:2012tn}. This suggests that the leading uncertainties in the
cross section for neutrino capture are the uncertainties of the matrix
elements of nuclear transitions to excited states. The analysis of the
capture cross sections for all three types of neutrino sources
considered in this paper reveals that taking into account all the
uncertainties in the determination of the threshold energy of the
gallium transition to the ground state of germanium and taking into
account the uncertainties of the matrix elements of the transitions to
excited states give an uncertainty of the cross sections of about
2\,\%. This result shows that the central values and errors of the cross sections
\eqref{Chrome-cross}-\eqref{Zinc-cross} cannot explain the 
anomalous results of  SAGE \cite{Abdurashitov:1998ne,Abdurashitov:2005tb}
 and GALLEX \cite{Kaether:2010ag}: the anomalous results remain intact. 

Thus, the main results published in \cite {Barinov:2016znv} where
the data \cite{PhysRevC.91.034608},
\cite{Frekers2013233} have not used, remain true, and the experiment BEST
\cite{Gavrin:2015aca} has high potential in testing the
hypothesis of electron neutrino oscillations into sterile neutrinos.

\begin{figure}[htb!]
	\centering
	\includegraphics[width=0.9\linewidth,
          height=0.3\textheight]{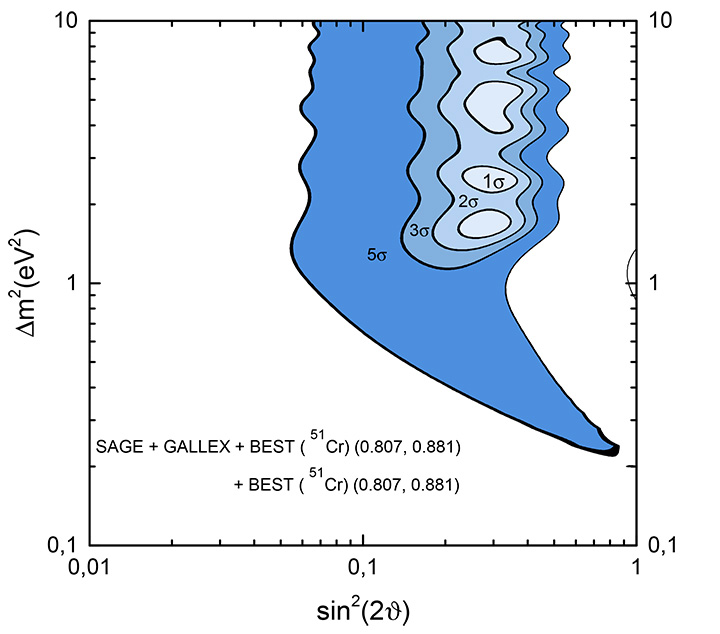}
	\caption{Allowed regions of oscillation parameters
		in case the BEST experiment confirms the gallium
                anomaly in both runs performed with the chrome-51 sources.
		\label{fig:SG51CrBB_bf}}
\end{figure}

To summarize, we present the refined estimates of BEST sensitivity to
models with light sterile neutrinos mixed with electron neutrinos. The
obtained results strongly 
suggest to use the new artificial source based on the isotope
$^{65}\!\Zn$ at the second stage of BEST operation, which allow us
to reduce the degeneracy in sensitivity to the sterile neutrino model
parameters. To illustrate this point we present in
  Fig.\,\ref{fig:SG51CrBB_bf}   
    the sensitivity contours in case of both stages exploiting
      the $^{51}\!\Cr$ sources. One can conclude by comparing the
      plots in
      Figs.\,\ref{fig:SGB1B2_BF_NS_NR},\,\ref{fig:SG51CrBB_bf} that
      while 5-$\sigma$ discovery of the sterile neutrinos is mostly
      due to double statistics (one stage is not enough to achieve this
      goal), the second source with different neutrino energies
      definitely provides with better cornering the signal regions
      with respect to the case of identical sources. 
We study possible impact of the future BEST results on 
status of the Gallium anomaly.

\vskip 0.3cm
We thank S.\,Kulagin, F.\,Simkovic and O.\,Smirnov for valuable
discussions. 
The work was supported by the RSF grant 17-12-01547.

\bibliographystyle{apsrev4-1}
\bibliography{Zinc-rev}

\end{document}